\documentclass{article}

\PassOptionsToPackage{numbers,compress}{natbib}

\usepackage[preprint]{neurips_2021}

\usepackage[utf8]{inputenc} 
\usepackage[T1]{fontenc}    
\usepackage{hyperref}       
\usepackage{url}            
\usepackage{booktabs}       
\usepackage{amsfonts}       
\usepackage{nicefrac}       
\usepackage{microtype}      
\usepackage{xcolor}         
\usepackage{xspace}
\usepackage{booktabs}
\usepackage{subfigure}
\usepackage{graphicx}
\usepackage{amsmath}
\usepackage{multirow}
\usepackage{bm}
\title{Dual-view Molecule Pre-training}

\author{%
  Jinhua Zhu $^1$, Yingce Xia $^2$, Tao Qin $^2$, Wengang Zhou $^1$, Houqiang Li $^1$, Tie{-}Yan Liu $^2$\\
  $^1$ University of Science and Technology of China\quad$^2$Microsoft Research Asia\\
  \texttt{teslazhu@mail.ustc.edu.cn,\;\{zhwg,lihq\}@ustc.edu.cn} \\
  \texttt{\{yinxia,taoqin,tyliu\}@microsoft.com}
}

\newcommand{\ourM}{DMP}

\newcommand{\ourMTF}{DMP$_\textrm{TF}$}
\newcommand{\ourMGNN}{DMP$_\textrm{GNN}$}

\begin{document}

\maketitle

\begin{abstract}
Inspired by its success in natural language processing and computer vision, pre-training has attracted substantial attention in cheminformatics and bioinformatics, especially for molecule based tasks. A molecule can be represented by either a graph (where atoms are connected by bonds) or a SMILES sequence (where depth-first-search is applied to the molecular graph with specific rules). Existing works on molecule pre-training use either graph representations only or SMILES representations only. In this work, we propose to leverage both the representations and design a new pre-training algorithm, dual-view molecule pre-training (briefly, \ourM{}), that can effectively combine the strengths of both types of molecule representations. The model of \ourM{} consists of two branches: a Transformer branch that takes the SMILES sequence of a molecule as input, and a GNN branch that takes a molecular graph as input. The training of \ourM{} contains three tasks: (1) predicting masked tokens in a SMILES sequence by the Transformer branch, (2) predicting masked atoms in a molecular graph by the GNN branch, and (3) maximizing the consistency between the two high-level representations output by the Transformer and GNN branches separately. After pre-training, we can use either the Transformer branch (this one is recommended according to empirical results), the GNN branch, or both for downstream tasks. \ourM{} is tested on nine molecular property prediction tasks and achieves state-of-the-art performances on seven of them. Furthermore, we test \ourM{} on three retrosynthesis tasks and achieve state-of-the-art results on them. 
\end{abstract}

\section{Introduction}
While machine learning (especially deep learning) techniques have been applied to cheminformatics and bioinformatics with significant progress~\cite{huang2016communication,david2020molecular}, their potential is severely limited by the scale of labeled data since it is more costly and time consuming to collect labeled data for tasks in cheminformatics and bioinformatics than those tasks in computer vision (CV) and natural language processing (NLP). Inspired by the great success of exploiting unlabeled data in CV and NLP  \cite{he2020momentum,grill2020bootstrap}, pre-training has been introduced into cheminformatics~\cite{hu2019strategies,chithrananda2020chemberta,wang2021molclr} and bioinformatics~\cite{rao2019evaluating,ESMpaper}. Among those  pre-training works, molecule pre-training has attracted much attention since molecules are a kind of basic units and play a fundamental  role in drug discovery and chemical modeling.

Molecules can be represented in different formats. As shown in Figure~\ref{fig:framework}, a molecule can be represented by a molecular graph, where nodes and edges are atoms and bonds respectively. This is the most intuitive way to visualize a molecule. By traversing the graph using depth first search and some predefined rules, the same molecule can be represented by a SMILES\footnote{SMILES is short for simplified molecular-input line-entry system.} \cite{weininger1988smiles} sequence. In almost all molecular database like PubChem~\cite{10.1093/nar/gkaa971} and ZINC~\cite{doi:10.1021/ci049714}, SMILES representations are used due to its simplicity. Correspondingly, there are mainly two lines of works on molecule pre-training, which are built upon graphs or  SMILES sequences. For pre-training on graphs \citep{hu2019strategies,liu2019ngram,wang2021molclr}, graph neural networks (GNNs) are used as backbone models, which explicitly leverages the structural information of a molecule. \citet{hu2019strategies} proposed two pre-training techniques based on GNNs: one is to predict masked atoms or bonds, and the other is to preserve the similarity between a $K$-hop subgraph of some anchor point $v_a$ and its corresponding context graph (i.e., a ring surrounding $v_a$). \citet{wang2021molclr} applied contrastive learning~\cite{1640964} to molecule pre-training, where the representations of a molecule extracted by a GNN are forced to be similar to another augmented version of itself while dissimilar to other molecules. For pre-training on SMILES sequences \citep{chithrananda2020chemberta}, they are treated in a similar way to natural language sequence and Transformer~\cite{vaswani2017attention}, a widely used model in NLP, is adopted as the backbone.. 

\begin{figure}[!ht]
    \centering
    \includegraphics[width=0.9\linewidth]{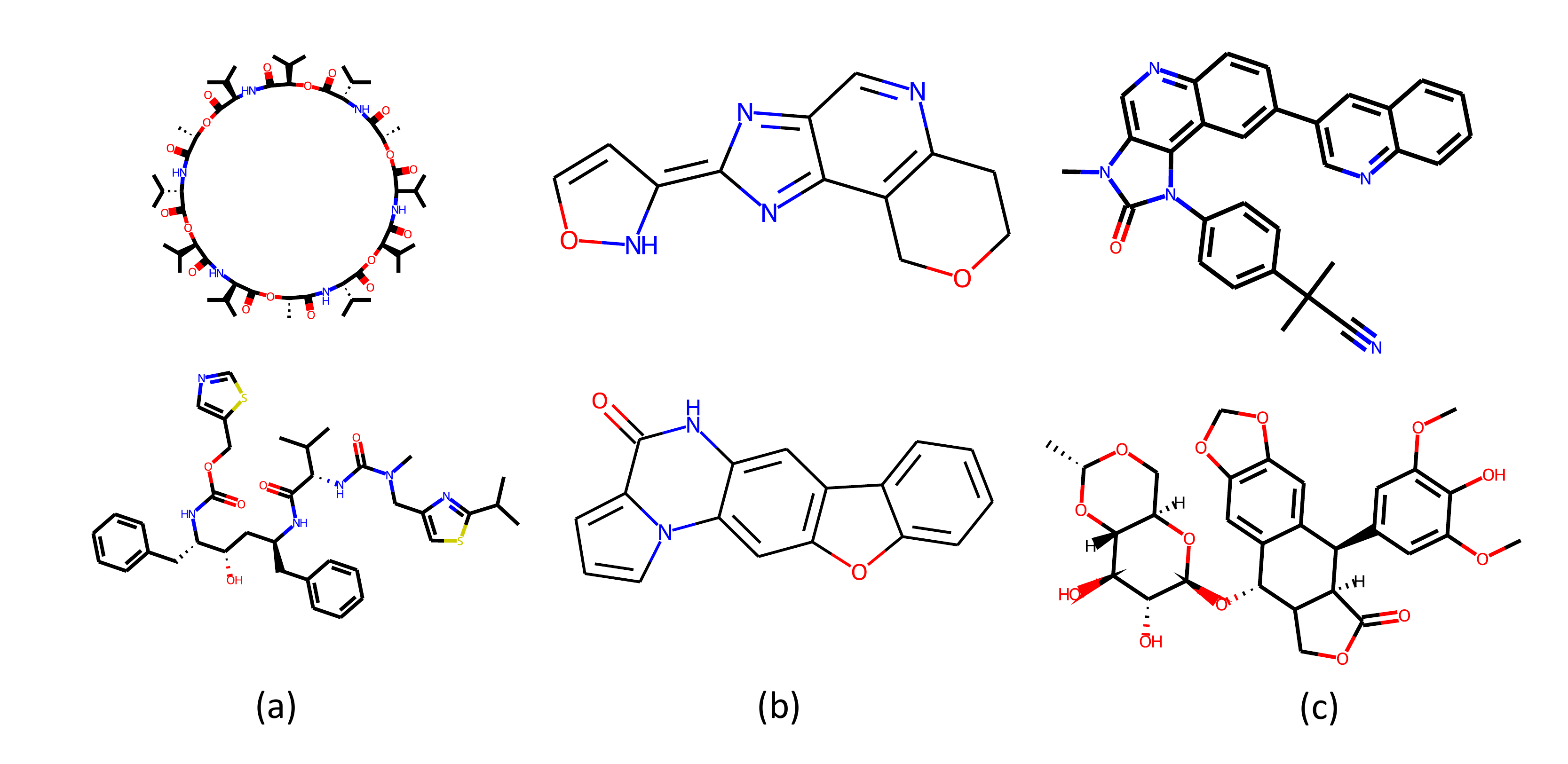}
    \caption{Examples of molecular property prediction from MoleculeNet~\cite{wu2018moleculenet}. (a) Transformer succeeds while GNN fails; (b) GNN succeeds while Transformer fails; (c) both the standard Transformer and GNN fail while our method succeeds. }
    \label{fig:case_study_in_motivation}
\end{figure}

Although different methods have been proposed for molecule pre-training, to our best knowledge, almost all of them only use one kind of molecule representations, either dealing with graphs using GNN only or dealing with SMILES sequences using Transformer only. However, the two types of models have their own strengths and limitations, as shown in Figure~\ref{fig:case_study_in_motivation}. Transformer correctly predicts the properties of molecules in Figure~\ref{fig:case_study_in_motivation}(a), whose maximal node/atom distance is large, but fails on the molecules in Figure~\ref{fig:case_study_in_motivation}(b) which has more than three rings concatenated together.
In contrast, GNN correctly classifies the molecules in Figure~\ref{fig:case_study_in_motivation}(b) but fails in Figure~\ref{fig:case_study_in_motivation}(a). This suggests that the two views are complementary to each other.

In this work, we propose a novel pre-training method, {\it dual-view molecule pre-training} (briefly, \ourM{}), to combine the best of two worlds. In \ourM{}, a molecule $M$ is represented by both a SMILES sequence $M_s$ and a graph $M_g$. $M_s$ is encoded by a Transformer branch which outputs a high-level representation $f_s$, and $M_g$ is encoded by a GNN branch which outputs another high-level representation $f_g$. Since $f_s$ and $f_g$ are representations of the same molecule, they should be similar in some latent space. To achieve this, inspired by the BYOL scheme~\cite{BYOL2020}, we propose the {\it dual-view consistency objective}, where the cosine similarity between the projected variants $f_s$ and $f_g$ should be maximized. In addition, we also adopt the masked language modeling (MLM)~\cite{hu2019strategies,chithrananda2020chemberta,devlin2018bert} objective. Specifically, for the Transformer branch, we randomly mask some tokens (e.g., atoms or bonds) in a SMILES sequence and recover them; for the GNN branch, we randomly mask some atoms and reconstruct them. After pre-training, we can use the Transformer branch, the GNN branch or both for downstream tasks. We recommend using the Transformer branch according to empirical results, which do not introduce extra parameters for downstream tasks compared to standard Transformer while the accuracy is promising. 

To test \ourM{}, we first pre-train on $10M$ molecules from PubChem\footnote{\url{https://pubchem.ncbi.nlm.nih.gov/}} following previous work~\cite{chithrananda2020chemberta,wang2021molclr}, and then finetune on $9$ molecular property prediction tasks from MoleculeNet \citep{wu2018moleculenet} and three retrosynthesis tasks. We achieve state-of-the-art results on $7$ out of $9$ prediction tasks, which demonstrate the effectiveness of our algorithm. Specifically, on the classification tasks of MoleculeNet, \ourM{} outperforms MolCLR~\cite{wang2021molclr} and GROVER~\cite{rong2020self} by $1.2$ and $2.2$ points on average; on retrosynthesis, we  achieve $2\sim 3$-point improvements of top-1 accuracy across 3 different settings and achieve state-of-the-art results on them \citep{GLN,NEURIPS2020_819f46e5,tetko2020state}. After that, we conduct pre-training on $100M$ compounds from PubChem, and we find the results can be slightly improved. After using our method, we can successfully predict the properties 
of molecules with long chains and rich structured groups (as shown in Figure~\ref{fig:case_study_in_motivation}(c)), where both standard Transformer only and GNN fail.

Our main contributions can be summarized as follows:

\noindent(1) To our best knowledge, we are the first to conduct molecule pre-training taking the advantages of the two different views (i.e., SMILES and molecular graphs).

\noindent(2) In addition to MLM, \ourM{}~leverages dual-view consistency loss for pre-training, which explicitly exploits the consistency of representations between two views of molecules.

\noindent(3) We achieve state-of-the-art results on $7$ molecular property prediction tasks from MoleculeNet~\cite{wu2018moleculenet} and three retrosynthesis tasks (i.e., a kind of molecule generation task), demonstrating the effectiveness and generalization ability of \ourM{}.

\section{Related work}

In this section, we briefly summarize molecule pre-training based on Transformer and GNN. 

\noindent{\bf Transformer-based pre-training}: Transformer~\cite{vaswani2017attention} is originally proposed for sequence learning in NLP, and then widely adopted by other applications, such as speech processing \cite{8462506,NEURIPS2019_f63f65b5}, CV \citep{dosovitskiy2021an,image2018imagetransformer}, and also, cheminformatics. To leverage Transformer-based molecule pre-training, people regard SMILES sequences as natural languages and apply the techniques of pre-training in NLP. Similar to~\cite{devlin2018bert,liu2019roberta} which mainly uses masked language modeling, \citet{wang2019smiles} proposed SMILES-BERT and \citet{chithrananda2020chemberta} ChemBERTa, which is to apply masked language modeling to molecule pre-training. 
\citet{fabian2020molecular} introduced additional tasks to SMILES-based molecule pre-training, like predicting whether two SMILES represent the same molecule, or predicting whether some molecular descriptors obtained by RDKit exist in a molecule.
\citet{honda2019smiles} used auto-encoder to obtain a neural-based fingerprint. \citet{maziarka2020molecule} improved self-attention with inter-atomic distances and molecular graph structure. In addition, there are some works about SMILES-based molecule generation \citep{schwaller2019molecular,pesciullesi2020transfer} and optimization \citep{he2021transformer}. 

\noindent{\bf GNN-based pre-training}: \citet{hu2019strategies} used a GNN to encode the input molecule, and proposed two pre-training strategies, where we should either recover the masked attributes of the input (e.g., atom type), or use constrastive learning~\cite{1640964,chen2020simple} to minimize the difference between two subgraphs of within a molecule. A similar idea also exists in \citet{li2020learn}. \citet{wang2021molclr} applied contrastive learning across different molecules and proposed MolCLR, where a molecule should be similar to an augmented version of itself while dissimilar to others. \citet{liu2019ngram} proposed an $N$-gram graph to represent molecules: it first learns the representations of atoms using CBoW~\cite{mikolov2013efficient}, and then enumerates $N$-grams (i.e., a walk of length $N$) in the graph and the final representation is constructed based on the embeddings of all its $N$-grams. \citet{shen2020molgnn} leveraged both edge-level reconstruction and motif-level reconstruction. \citet{rong2020self} proposed GROVER, which replaces the self-attention layer in Transformer where the number of hops are adaptively determined, and design new pre-training objective functions: one about predicting the statistics of the molecule and the other is the predict the existence of some function groups. To our best knowledge, almost all previous works treat Transformer-based pre-training and GNN-based pre-training independently, and we propose to leverage them together.

\section{Our method}
In this section, we first introduce the network architecture, followed by the training objective functions of our method, and finally provide discussions about related methods and limitations.

\subsection{Network architecture}
\begin{figure}[!ht]
    \centering
    \includegraphics[width=0.95\linewidth]{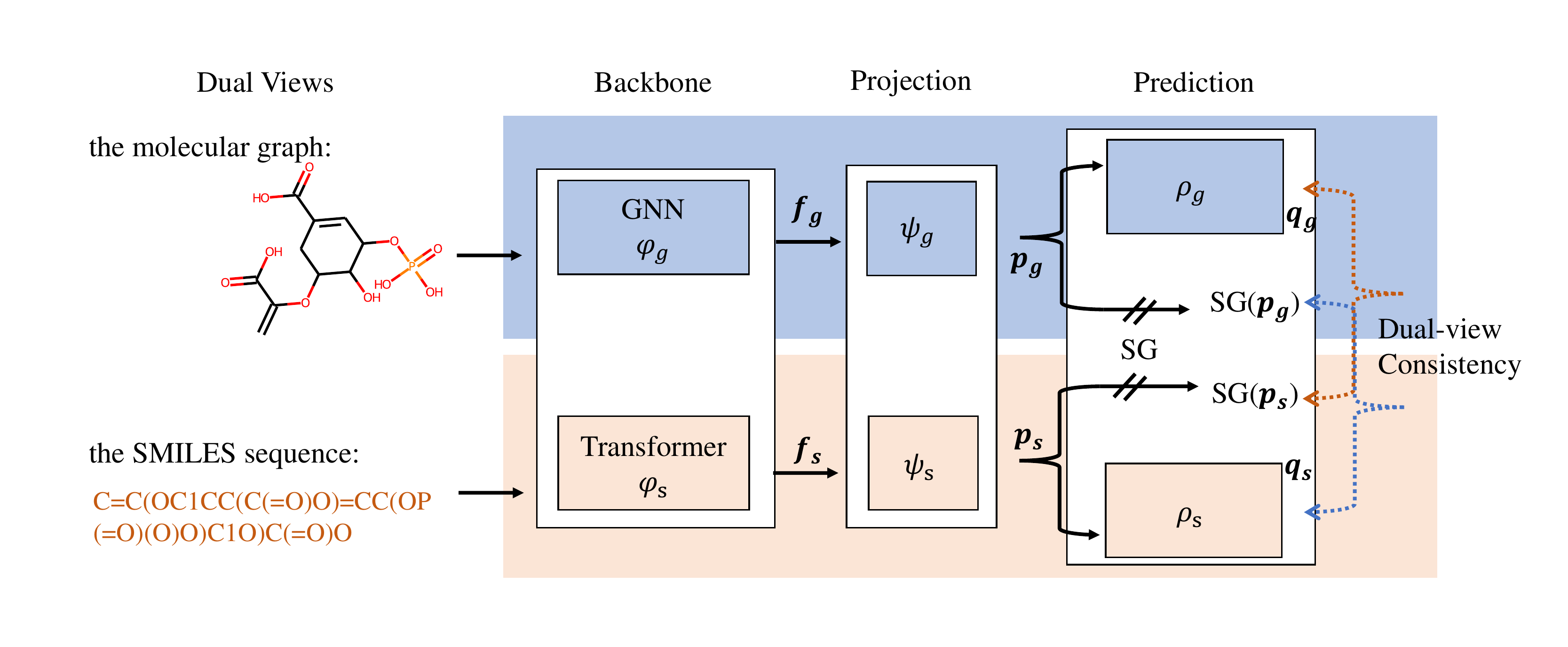}
    \caption{The illustration of our dual-view consistency loss. "SG" indicates the stop-gradient when back-propagating.}
    \label{fig:framework}
\end{figure}

Given a molecule $M$, let $M_s$ and $M_g$ denote the SMILES and molecular graph respectively. Define $M_s$ as $(m_1,m_2,\cdots,m_l)$, where $l$ is the length of $M_s$.
Define $M_g$ as a graph $(V_g, E_g)$, where $V_g=\{v_1,v_2,\cdots,v_n\}$ is a collection of atoms, and $E_g$ is the collection of edges\footnote{$l\geq n$ since the bonds, numbers and brackets are also included in the SMILES as shown in Figure~\ref{fig:framework}.}. In a molecule, the bond has several types (e.g., single bond, double bonds, aromatic compounds), which is also modeled. 

The network architecture of our method consists of two branches, one is a GNN, which is used to encode its graph view $M_g$, and the other one is a Transformer, which is to encode its sequence view $M_s$. Denote the above two models as $\varphi_g$ and $\varphi_s$, which have $L_g$ and $L_s$ layers, respectively.

\noindent(1) For the GNN branch, following~\cite{li2020deepergcn}, we choose the DeeperGCN network as the backbone. DeeperGCN is a stack of multiple GCN layers~\cite{kipf2016semi}, where the batch normalization~\cite{ioffe2015batch}, non-linear activation and residual connections~\cite{he2016deep} are all used. The information among atoms are propagated along bonds, and the embedding of bond property (i.e., double bond, single bond) is added to atoms. Details of DeeperGCN are left in Appendix \ref{appendix:deepergcn} of the supplementary document. After encoded by the $L_g$ layers, each atom has a representation outputted by the last layer of DeeperGCN, i.e., $\{h^g_1,h^g_2,\cdots,h^g_n\}=\varphi_g(M_g)$, where $h^g_i$ is the representation of the $i$-th atom. We then choose a pooling function $\operatorname{pool}$ (e.g., mean pooling, max pooling) and obtain the representation of the molecule, i.e., $f_g=\operatorname{pool}(h^g_1,h^g_2,\cdots,h^g_n)$. 

\noindent(2) For the Transformer branch, we choose the RoBERTa model~\cite{liu2019roberta}, which encodes the SMILES sequence $M_s$ of the molecule. Following \cite{liu2019roberta,devlin2018bert}, we add a special token \texttt{[CLS]} to the beginning of the sequence. We also use the output of the last layer as the representations of \texttt{[CLS]} and tokens, i.e., $\{h^s_0,h^s_1,h^s_2,\cdots,h^s_l\}=\varphi_s(M_s)$, where $h^s_0$ corresponds to \texttt{[CLS]} and $h^s_j$ corresponds to the $j$-th element in the SMILES sequence. In this way, the molecule representation of $M_s$ is $h^s_0$, i.e., $f_s=h^s_0$.

\subsection{Training objective functions}
Our method consists of three training objective functions, including two masked language modeling (MLM) loss and one dual-view consistency loss.

\noindent(1) {\bf MLM on Transformer}: Given a SMILES sequence, we randomly mask some tokens, and the objective is to recover  the original tokens, following the practice in NLP pre-training~\cite{devlin2018bert,liu2019roberta}.

\noindent(2) {\bf MLM on GNN}: Similar to the MLM on Transformer, we randomly mask some atoms in a molecular graph (the bonds remain unchanged), and the objective is to recover the original atoms, following the practice in graph pre-training~\cite{hu2019strategies}.

\noindent(3) \textbf{Dual-view consistency}: To model the interaction between the GNN and the Transformer, inspired by the BYOL~\cite{BYOL2020}, we propose a dual-view consistency loss, that models the similarity of the output between GNN and Transformer. Based on the empirical discovery of~\cite{BYOL2020}, we introduce two non-linear projection layers $\psi_g$, $\psi_s$ and two prediction layers $\rho_g$ and $\rho_s$. For the SMILES view $M_s$ and 
the graph view $M_g$ of a molecule, we randomly mask some tokens/atoms and obtain $\tilde{M}_s$ and $\tilde{M}_g$.
$f_g$ is obtained by $\operatorname{pool}(\varphi_g(\tilde{M}_g))$ and $f_s$ is the first output of $\varphi_s(\tilde{M}_s)$, which corresponds to the \texttt{[CLS]} token.
After that, we  apply the projection and prediction layers to them, i.e., 
\begin{equation}
p_g = \psi_g(f_g),\;q_g=\rho_g(p_g);\;p_s = \psi_s(f_s),\;q_s=\rho_s(p_s).
\end{equation}
Since $p_{\cdot}$'s and $q_{\cdot}$'s are the representation of the same molecule but from different views, they should preserve enough similarity. Let $\texttt{cos}(p,q)$ denote the cosine similarity between $p$ and $q$, i.e., $\texttt{cos}(p,q)=(p^\top q)/(\Vert p\Vert_2\Vert q\Vert_2)$. Following~\cite{BYOL2020}, the dual-view consistency loss is defined as 
\begin{equation}
    \ell_{\rm dual}(\tilde{M}_g,\tilde{M}_s;\varphi_g,\varphi_s,\psi_g, \psi_s, \rho_g, \rho_s) = -\texttt{cos}(q_s, \operatorname{SG}(p_g)) - \texttt{cos}(q_g, \operatorname{SG}(p_s)).
    \label{eq:dual_view_loss}
\end{equation}
In Eqn.\eqref{eq:dual_view_loss}, $\operatorname{SG}(p_g)$ means that the gradient is not applied to $p_g$ when back-propagating, and neither is $\operatorname{SG}(p_s)$.

After pre-training, we can use the Transformer branch, the GNN branch or both for downstream tasks. According to our empirical study, for molecular prediction tasks, we recommend using the Transformer branch. Using both branches brings further improvement at the cost of larger model size.

\subsection{Discussions}
(1) {\it Relation with co-training}: Considering there are two ``views'' in our method, people might be curious the the relation with co-training, which also leveraged two views. Co-training~\cite{blum1998combining} is a semi-supervised learning method that can be used to construct additional labeled data. The basic assumption to use co-training is that the features can be divided into two conditionally independent sets (i.e., views), where each set is sufficient to train a classifier. Each classifier provides the most confident predictions on the unlabeled data as the additional labeled data. 
In comparison, \ourM{} is an algorithm which does not need the independent assumption. \ourM{} provides a pre-trained model for downstream tasks instead of newly labeled data.

In addition, there are some work leveraging multiple views for pre-training, but the views are usually from the same model instead of two heterogeneous ones (Transformer vs GNN). In \citep{ma2020multi}, the two views are the outputs of a node-central encoder and an edge-central encode. In \citep{hassani2020contrastive}, the two views of a graph are the first-order neighbors
and a graph diffusion. Those methods are significantly different from ours.

(2) {\it Limitations of our method}:\label{discussion_limitation} Our method has two limitations. First, there is a Transformer branch and a GNN branch in our method, which increases the training cost comparing with previous single-branch pre-training~\cite{chithrananda2020chemberta,wang2021molclr}. How to design an efficient pre-training method is an interesting future direction. Second, in downstream tasks, we deal with all molecules using either the Transformer branch or the GNN branch. Recent studies show that a better solution is to use a meta-controller to dynamically determine which branch to use~\cite{zhu2021iot,elbayad2019depth} for an individual input. We will also explore this dynamic branch selection in future. 


\section{Experiments}
In this section, first, we introduce the data and model architecture for pre-training. Second, we apply our pre-trained models to nine molecular property prediction tasks from MoleculeNet~\cite{wu2018moleculenet}. Next, we evaluate our method on three retrosynthesis tasks, including USPTO-50k with reaction type known/unknown and USPTO-full~\cite{tetko2020state}. Finally, we visualize the representations of different pre-trained models for comparison.
\subsection{Pre-training}
\noindent\textbf{Dataset}
For pre-training, we choose two subsets from PubChem, one with $10M$ compounds which are the same as those in \cite{wang2021molclr,chithrananda2020chemberta}, and the other with $100M$ compounds. We first use the RDkit toolkit\footnote{\url{https://github.com/rdkit/rdkit}} to process the input. For the input of Transformer branch, we canonicalize SMILES sequences and then tokenize the canonicalized SMILES using the same regular expression as \cite{schwaller2019molecular}. For the input of the GNN branch, we convert the SMILES to molecular graphs using RDKit. 

\noindent\textbf{Model Architecture}
The Transformer branch is the same as the RoBERTa$_{\rm base}$ architecture, which consists of $12$  layers. The hidden dimension, the dimension of the feed-forward layer, and the number of heads are $768$, $3072$ and $12$, respectively. The GNN branch follows the DeeperGCN \citep{li2020deepergcn} backbone, which is a $12$-layer network. The hidden dimension of the GNN branch is $384$. The graph pooling function above the last layer is the concatenation of \texttt{mean} and \texttt{max} operations. 

Following \cite{chen2020simple, BYOL2020}, the projection heads (i.e., $\psi_g$ and $\psi_s$) are 3-layer MLP networks, and the prediction heads (i.e., $\rho_g$ and $\rho_s$) are 2-layer MLP networks. All hidden layers in the MLPs are followed by the \texttt{ReLU} and \texttt{BatchNorm} \citep{ioffe2015batch}, while the output layers are not.

\noindent\textbf{Optimization}
Our model is optimized by Adam \citep{kingma2014adam} algorithm with learning rate $5\times10^{-4}$, $\beta_1=0.9$, $\beta_2=0.98$, $\epsilon=10^{-6}$. The weight decay is $0.01$. The learning rate is warmed up in the first $10k$ update steps and then linearly decayed. All pre-training experiments are conducted on $8\times$V100 GPUs and the batch size is $12288$ tokens. The  gradients are accumulated for $16$ times. The models are trained for $200k$ iterations. 

\subsection{Molecular property prediction}
\noindent\textbf{Dataset} 
After pre-training, we first finetune our model on 6 datasets from MoleculeNet~\cite{wu2018moleculenet}, a popular well-curated benchmark for molecular property prediction. Note that each dataset might be associated with multiple tasks. There are $44$ tasks in total, and they cover a wide range of data volumes. On one hand, following \cite{wang2021molclr}, we use the official training, validation and test sets provided by DeepChem~\footnote{\url{https://github.com/deepchem/deepchem}}, whose performance is relatively stable and reproducible.
On the other hand, to compare with  \citep{rong2020self,li2020learn}, we follow their ways to split the data and conduct another group of experiments. Specifically, we use {\it scaffold splitter} to generate the training ($80\%$), validation ($10\%$) and test ($10\%$) sets with 3 different seeds. We compare with \citep{rong2020self,li2020learn} on three classification and three regression tasks. 

\noindent\textbf{Finetuning settings} Following \cite{liu2019ngram,wang2021molclr}, each task is independently finetuned. We explore finetuning on the Transformer branch, the GNN branch and both. The classification/regression head is a $2$-layer MLP with \texttt{tanh} activation, which takes the representation of \texttt{[CLS]} (i.e., $f_s$) and/or the graph pooling over GNN (i.e., $f_g$) as input. Detailed hyper-parameters are left in Appendix \ref{appendix:finetuninghyperparaments}.

\noindent\textbf{Evaluation}
Following \cite{wu2018moleculenet}, for classification tasks, we use the area under curve (AUC) of the receiver operating characteristic (ROC) curve as the evaluation metric. For regression tasks, we use the root mean square error (RMSE) or the mean absolute error (MAE) for evaluation depending on the previous work. For datasets containing more than one task, we report the average scores across all tasks. Each task is independently run for 3 times with different random seeds, and the mean and standard derivation of the performances are reported. 

\noindent\textbf{Baselines} We compare our method with the following baselines:

\noindent(1) We compare \ourM{} with four methods without pre-training. Two of them are about applying Random Forest (RF) \citep{ho1995random} and Support Vector Machine (SVM) \citep{cortes1995support} to molecular fingerprints. The other two are GNN-based method, D-MPNN \citep{yang2019analyzing} and MGCNN \citep{lu2019molecular}, which are specifically designed for molecular property prediction. We also compare our model with previous pre-training methods, including \citet{hu2019strategies}, MolCLR~\citep{wang2021molclr}, GROVER \citep{rong2020self} and MPG \citep{li2020learn}.


\noindent(2) We train a standard Transformer and GNN using MLM only, which are denoted as TF (MLM) and GNN (MLM). TF (MLM) is an enhanced implementation of ChemBERTa~\cite{chithrananda2020chemberta}, where we train models with more GPUs and larger batch size, and eventually obtain better results (see Appendix \ref{appendix:chemberta} for more details). For GNN (MLM), we only mask the atoms randomly, without applying bond deletion and subgraph removal, which are left as future work. Also, we implement a variant of \ourM{}, ``\ourM{} w/o MLM'',  where we only use dual-view loss and do not use MLM.

\noindent(3) To investigate the effectiveness of using two heterogeneous models, we implement another variants, where the two branches are both Transformer or both GNNs. The MLM loss and dual-view consistency loss are both applied. We randomly choose one branch for finetuning. Denote these variants as TF ($\times2$) and GNN ($\times2$).

\begin{table}[!htb]
    \centering
    
    \resizebox{1\textwidth}{!}{
    \begin{tabular}{lcccccc}
    \toprule
 Dataset & BBBP & Tox21 & ClinTox & HIV & BACE & SIDER  \\
\# Molecules & 2039 & 7831 & 1478 & 41127 & 1513 & 1478  \\
\midrule
RF & $71.4 \pm 0.0$ & $76.9 \pm 1.5$ & $71.3 \pm 5.6$ & $78.1 \pm 0.6$ & $86.7 \pm 0.8$ & $68.4 \pm 0.9$  \\
SVM & $72.9 \pm 0.0$ & $81.8 \pm 1.0$ & $66.9 \pm 9.2$ & $79.2 \pm 0.0$ & $86.2 \pm 0.0$ & $68.2 \pm 1.3$  \\
MGCN \citep{lu2019molecular} & $\bm{85.0 \pm 6.4}$ & $70.7 \pm 1.6$ & $63.4 \pm 4.2$ & $73.8 \pm 1.6$ & $73.4 \pm 3.0$ & $55.2 \pm 1.8$  \\
D-MPNN \citep{yang2019analyzing} & $71.2 \pm 3.8$ & $68.9 \pm 1.3$ & $90.5\pm5.3$ & $75.0 \pm 2.1$ & $85.3 \pm 5.3$ & $63.2 \pm 2.3$  \\
\citet{hu2019strategies} & $70.8 \pm 1.5$ & $78.7 \pm 0.4$ & $78.9 \pm 2.4$ & $80.2 \pm 0.9$ & $85.9 \pm 0.8$ & $65.2 \pm 0.9$ \\
MolCLR~\cite{wang2021molclr} & $73.6 \pm 0.5$ & $\bm{79.8\pm0.7}$ & $93.2\pm1.7$ & $80.6 \pm 1.1$ & $89.0\pm0.3$ & $68 .0 \pm1.1$  \\
\midrule
TF (MLM)& $74.9\pm0.6$&$77.6\pm0.4$&$92.9\pm0.5$& $80.2\pm0.4$&$88.0\pm0.5$&$68.4\pm0.4$\\
TF ($\times2$)& $75.6\pm0.7$ & $77.1\pm0.5$&$92.0\pm0.8$&$80.4\pm0.4$&$88.1\pm0.5$ & $68.2\pm1.2$\\
\ourMTF{} w/o MLM  & $71.1\pm0.4$ &$75.7\pm0.4$&$93.8\pm0.7$&$79.1\pm1.7$&$88.3\pm0.7$&$68.1\pm0.7$\\
\ourMTF{} &$78.1\pm0.5$ &$78.8\pm0.5$ &$\bm{95.0\pm0.5}$&$\bm{81.0\pm0.7}$&$\bm{89.3\pm0.9}$&$\bm{69.2\pm0.7}$ \\
\midrule
GNN (MLM) & $74.5\pm0.3$&$74.8\pm0.5$&$92.3\pm0.7$&$78.5\pm0.5$&$84.1\pm0.4$&$67.0\pm0.5$\\
GNN ($\times2$) & $74.1\pm0.6$&$75.1\pm0.3$ & $92.8\pm0.7$&$79.2\pm0.9$& $85.1\pm1.0$&$69.0\pm0.4$\\
\ourMGNN{} &$74.7\pm0.2$&$76.7\pm0.3$&$94.2\pm0.4$&$79.5\pm1.0$&$85.7\pm0.8$&$68.4\pm0.5$ \\
\midrule
TF (MLM) + GNN (MLM) & $76.1\pm0.3$ & $77.8\pm0.8$ & $94.0\pm0.4$ & $80.1\pm0.4$ & $87.5\pm0.9$ & $69.3\pm0.9$\\
\ourM{}$_\textrm{TF+GNN}$&$77.8\pm 0.3$& $79.1\pm0.4$&$95.6\pm0.7$&$81.4\pm0.4$&$89.4\pm0.8$&$69.8\pm0.6$\\
\ourMTF{} (100M) &$78.4\pm0.3$&$79.0\pm0.3$&$95.5\pm0.2$&$81.1\pm0.3$&$89.6\pm0.3$&$70.0\pm0.6$ \\
\ourMGNN{} (100M)&$75.2\pm0.6$&$77.5\pm0.6$&$94.7\pm0.4$&$80.3\pm0.5$&$86.3\pm1.0$&$69.2\pm0.5$\\ 
\bottomrule
\end{tabular}}
    \caption{\small Test ROC-AUC (\%) performance of different methods on 6 binary classification tasks from MoleculeNet benchmark. The training, validation and test sets are provided by DeepChem in advance. The first parts are cited from \citet{wang2021molclr}. Each experiment is independently run for three times (with random seeds 0, 1, 2). The mean and standard derivation are reported.}
   
    \label{tab:results_overall_molecular_property_prediction}

\end{table}

\begin{table}[!htb]
\vspace{-0.2cm}
    \centering
    \resizebox{1\textwidth}{!}{
    \begin{tabular}{lccccccc}
    \toprule
          & \multicolumn{3}{c}{Classification (Higher is better)}&&\multicolumn{3}{c}{Regression (Lower is better)} \\
       \cmidrule{2-4}\cmidrule{6-8}
      Dataset &BBBP&SIDER&ClinTox&&ESOL &QM7&QM8\\
      Metric & ROC-AUC& ROC-AUC& ROC-AUC&&RMSE&MAE&MAE\\
       \midrule
       GROVER & ${0.940}_{(0.019)}$ & ${0.658}_{(0.023)}$ & ${0.944}_{(0.021)}$ &&{${0.831}_{(0.025)}$} & ${72.6}_{(3.8)}$ &  $0.0125_{(0.002)}$\\
       MPG& $0.922_{(0.012)}$ & ${0.661}_{(0.007)}$ & ${0.963}_{(0.028)}$ &&${0.741}_{(0.017)}$& $-$ & $-$\\
       \ourMTF{} & ${0.945}_{(0.020)}$ &${0.695}_{(0.011)}$&${0.968}_{(0.007)}$&&${0.700}_{(0.084)}$&${69.6}_{(8.3)}$&${0.0124}_{(0.002)}$\\
       \bottomrule
    \end{tabular}}
    \caption{\small The performance comparison on molecular property prediction. The raw data is from \cite{rong2020self}.
    }
    \label{tab:results_grover_mpg}

\end{table}

\noindent{\textbf{Results}} 
The results of the official test from DeepChem are shown in Table~\ref{tab:results_overall_molecular_property_prediction}, and those of the test sets from GROVER \cite{rong2020self} are in Table~\ref{tab:results_grover_mpg}. After pre-training with \ourM{}, denote the results finetuned from the Transformer branch and GNN branch as \ourMTF{} and \ourMGNN{} respectively. We have the following observations:

\noindent(1) Compared with the previous supervised methods, \ourMTF{} outperforms almost all previous baselines across different tasks, which leverage well-designed fingerprints or the specially designed GNNs. The results demonstrate  the effectiveness of using pre-trained models.

\noindent(2)  As shown in the second and third parts of Table~\ref{tab:results_overall_molecular_property_prediction}, both MLM loss and dual-view consistency loss help, no matter for the Transformer branch or GNN branch. Take the BBBP task of Transformer branch as an example. With MLM only or dual-view loss only, the accuracy is $74.9$ and $71.1$ respectively. After using both of them, the accuracy becomes $78.1$. On the other hand, if we apply dual-view loss only, we find that in general, the results are slightly worse than using MLM only (see the row ``\ourMTF{} w/o MLM'').

\noindent(3) If we apply our method to two Transformer branches or two GNN branches (i.e., the ``TF ($\times2$)'' and ``GNN ($\times2$)'' in Table~\ref{tab:results_overall_molecular_property_prediction}), we can see the results is not good as our proposed \ourM{}, although we observe some improvement over using MLM only or dual-view consistency loss only. This is consistent with our discovery in Figure~\ref{fig:case_study_in_motivation}, where the Transformer and GNN have complementary views towards processing molecules.

\noindent(4) We empirically found that tuning on the Transformer branch is better than tuning on the GNN branch. Therefore, by default, we recommend using the Transformer branch. However, if we do not count for the number of parameters and use both branches for inference, the performances can be further improved (see the row ``\ourM{}$_\textrm{TF+GNN}$''), but not too much.

\noindent(5) Compared with other pre-training methods, \citet{hu2019strategies} and MolCLR, which have more sophisticated models or training strategy, \ourMTF{} achieves the best performance on $5$ out of $6$ binary classification tasks. Also, compared with two recent models, GROVER and MPG,  \ourMTF{} also outperforms them on  classification and regression tasks (see Table \ref{tab:results_grover_mpg}). This shows that the joint pre-training of different views brings impressive benefit for molecular property prediction.

\noindent(6) We ensemble the TF (MLM) and GNN (MLM), where the two models are independently trained and finetuned, and their predictions are averaged. The results are denoted by the row ``TF (MLM) + GNN (MLM)''. Our method \ourMTF{} still significantly outperforms the ensemble baseline.
 
\noindent(7) Finally, we also pre-train on $100M$ compounds and then finetune on the downstream tasks. Compared to the results obtained by pre-training on $10M$ data, we observe improvement for both Transformer branch and GNN branch. We will explore how to effectively use more data in the future.

We compare the training time and parameters of different models. The statistics are reported in Table~\ref{tab:summary_modelpram_time}. DMP takes slightly longer training time than TF (MLM), (3.8 days v.s. 2.5 days), but achieves significant improvement over previous baselines.

\begin{table}[!htbp]
\centering
\small
\begin{tabular}{lll}
\toprule
& \# Parameters (M) & Training Time \\
\midrule
GROVER large~\cite{rong2020self} & 100  & 250 Nvidia V100 $\times$ 2.5 days \\
MPG & 53 & N/A \\
MolCLR~\cite{wang2021molclr} & N/A & 1 RTX 600 $\times$ 5 days \\
TF (MLM) & 87 & 8 Nvidia V100 $\times$ 2.5 days \\
TF ($\times 2$) & 184.6 & 8 Nvidia V100 $\times$ 5 days \\
GNN (MLM) & 7.4 & 8 Nvidia V100 $\times$ 1.7 days \\
GNN ($\times 2$) & 23.6 & 8 Nvidia V100 $\times$ 2.8 days \\
DMP & 104.1 & 8 Nvidia V100 $\times$ 3.8 days\\
\bottomrule
\end{tabular}
\caption{Statistics of model parameters and training time.}
\label{tab:summary_modelpram_time}
\end{table}

\subsection{Experiments on retrosynthesis}
In addition to molecule classification, we also conduct experiments on molecule generation. Specifically, we choose the retrosynthesis task: Given a target molecule (i.e., product) which cannot be directly obtained, we want to identify several easily obtained molecules (i.e., reactants) that can synthesize the product. 

\noindent{\bf Dataset}: Following~\citep{GLN,G2Gs,NEURIPS2020_819f46e5}, we conduct experiments on two widely used  datasets, USPTO-$50$K \citep{liu2017retrosynthetic,coley2017computer} and USPTO-full \citep{GLN,NEURIPS2020_819f46e5}. USPTO-$50$K consists of $50$K reactions with 10 reaction types in total, and USPTO-full consists of $950$K cleaned reactions from the USPTO 1976-2016 without reaction types. We use the data released by \cite{GLN}, where the training, validation and test has been split in advance and each part contains $80\%$, $10\%$ and $10\%$ of the total data respectively.
For USPTO-$50$K, we work on two settings where the reaction type is given or not.

\noindent{\bf Network Architecture}: We explore both the Transformer branch and the GNN branch.


For the Transformer branch, we implement three models for comparison: (i) the standard Transformer. Both the encoder and decoder have $6$ layers, with $4$ attention heads, embedding dimension $512$ and feed-forward dimension $1024$; (ii) We use the pre-trained model to initialize the encoder while the decoder remains the same as (i); (iii) Following~\cite{zhu2020incorporating}, we fuse the pre-trained branch with Transformer, where each layer in the encoder and decoder attends to the output of pre-trained model in an attentive way. 

For the GNN branch, we combine our method with GLN~\citep{GLN}. Specifically, we replace the GNN modules (i.e., the $g_1$ to $g_6$ in \citep{GLN}) used in its released code (\url{https://github.com/Hanjun-Dai/GLN}) by our pre-trained GNN branch, while keep the remaining part unchanged. Denote this method as ``GLN w/ DMP''.  We also implement another variant of GLN, where the GNN modules are replaced with DeeperGCN~\citep{li2020deepergcn} so as to verify the effectiveness of pre-training.



\noindent{\bf Evaluation Metrics}: We evaluate the models by the top-$k$ exact match accuracy (briefly, top-$k$ accuracy), which verifies that given a product, whether one of the $k$ generated reactant sets exactly matches the ground truth reactant set, $k\in\{1, 3, 5, 10, 20, 50\}$. For all $k>1$, we use beam search to generate the reactant sets, and rank them by log likelihoods.
\begin{table}[!htbp]
    \centering
    \small
    \begin{tabular}{lcccccc}
    \toprule
         \multirow{2}{*}{Methods} & \multicolumn{6}{c}{Top-$k$ accuracy (\%)}  \\
         \cmidrule{2-7}
         & $1$& $3$ & $5$& $10$&$20$ & $50$\\
         \midrule
         \multicolumn{7}{c}{Reaction types unknown on USPTO-50K}\\
         \midrule
         Transformer & $42.3$ & $61.9$ & $67.5$ & $72.9$ & $75.5$ & $77.1$ \\
         Pre-trained model as Encoder& $39.6$&$55.3$&$59.1$&$63.2$&$66.0$&$68.6$\\
         ChemBERTa \cite{chithrananda2020chemberta} fusion &$43.9$&$62.2$&$68.0$&$73.1$&$75.4$&$77.0$\\
         \ourM{} fusion &$46.1$&$65.2$&$70.4$&$74.3$&$76.1$&$77.5$ \\
         \midrule
         \multicolumn{7}{c}{Reaction types give as prior on USPTO-50K}\\
         \midrule
         Transformer & $54.2$ & $73.6$ & $78.3$&$81.3$& $83.1$&$84.3$ \\
         ChemBERTa fusion &$56.4$&$74.7$&$78.9$&$81.8$&$83.3$&$84.5$\\
          \ourM{} fusion &$57.5$&$75.5$&$80.2$&$83.1$&$84.2$&$85.1$\\
         \midrule
         \multicolumn{7}{c}{Retrosynthesis results on USPTO-full}\\
         \midrule
         Transformer & $42.9$ & $58.0$ & $62.4$ & $66.8$ &$69.8$& $72.5$ \\
         \ourM{} fusion &$45.0$&$59.6$&$63.9$&$67.9$&$70.7$&$73.2$ \\
         \bottomrule
    \end{tabular}
    \caption{Results of top-$k$ exact match accuracy on retrosynthesis (Transformer-based models).}
    \label{tab:retrosys}
\end{table}

\noindent{\bf Results of Transformer-based Models}: The results of the Transformer-based models are shown in Table \ref{tab:retrosys}. We first train the standard Transformer (denoted as ``Transformer''), which achieves $42.3\%$ top-1 accuracy on the unknown type setting. After initializing the encoder with pre-trained model, we observe that the accuracy drops to $39.6\%$. This is consistent with the discovery in \cite{zhu2020incorporating}, where simply applying initialization to sequence generation might not lead to good results. 

After using the method in \cite{zhu2020incorporating} (marked as \ourM{} fusion), we found that our model brings improvement to retrosynthesis. Specifically, on USPTO-50k dataset, we improve the top-1 accuracy from $42.3\%$ to $46.1\%$ when the reaction type is unknown, and from $54.2\%$ to $57.5\%$ when the type is given. On average, \ourM{} improves the standard Transformer by $2\sim 3$ points w.r.t. top-$1$ accuracy across all settings. Compared to a previous pre-trained model ChemBERTa~\cite{chithrananda2020chemberta} which is also used following~\cite{zhu2020incorporating}, our method outperforms it by $2.2$ and $1.1$ on the above two settings, which demonstrates the superiority of \ourM{} again. On the largest dataset, USPTO-full, our method also boosts the Transformer baseline by $2.1$ points, setting a state-of-the-art results on USPTO-full. The comparisons of our method with previous methods are shown in Appendix \ref{appendix:retrosys}.

\noindent{\bf Results of GNN-based Models}: The results of combining the GNN branch and GLN are shown in Table~\ref{tab:retrosys_gnn}. By combing GNN with our method, we achieved state-of-the-art results on these two tasks, which demonstrated the effectiveness of pre-training. By comparing with GLN w/ DeeperGCN, we can conclude that the improvement is brought by pre-training instead of different GNN modules.

\begin{table}[!htbp]
\centering
\small
\begin{tabular}{lcccccc}
\toprule
\multirow{2}{*}{Methods} & \multicolumn{6}{c}{Top-$k$ accuracy (\%)}  \\
\cmidrule{2-7}
& $1$& $3$ & $5$& $10$&$20$ & $50$\\
\midrule
\multicolumn{7}{c}{Reaction types unknown on USPTO-50K}\\
\midrule
GLN~\citep{GLN} & $52.5$ & $69.0$ & $75.6$ & $83.7$ & $89.0$ & $92.4$ \\
         GLN w/ DeeperGCN & $52.1$ & $68.8$&$76.1$&$84.0$&$89.2$&$92.4$\\
         GLN w/ \ourM{}  &$54.2$&$70.5$&$77.2$&$84.9$&$90.0$&$92.7$\\
\midrule
\multicolumn{7}{c}{Reaction types give as prior on USPTO-50K}\\
\midrule
GLN & $64.2$ & $79.1$ & $85.2$ & $90.0$ & $92.3$ & $93.2$ \\
GLN w/ DeeperGCN & $63.2$ & $79.2$&$85.0$&$90.0$&$92.0$&$93.1$\\
GLN w/ \ourM{}  &$66.5$&$81.2$&$86.6$&$90.5$&$92.8$&$93.5$\\
         \bottomrule
    \end{tabular}
    \caption{Results of top-$k$ exact match accuracy on retrosynthesis (GNN-based models).}
    \label{tab:retrosys_gnn}
\end{table}


\subsection{Visualization of pre-trained representations}
\begin{figure}[!htpb]
\centering
\begin{minipage}{0.33\linewidth}
\subfigure[GNN: DB Index 3.56]{
\includegraphics[width=\linewidth]{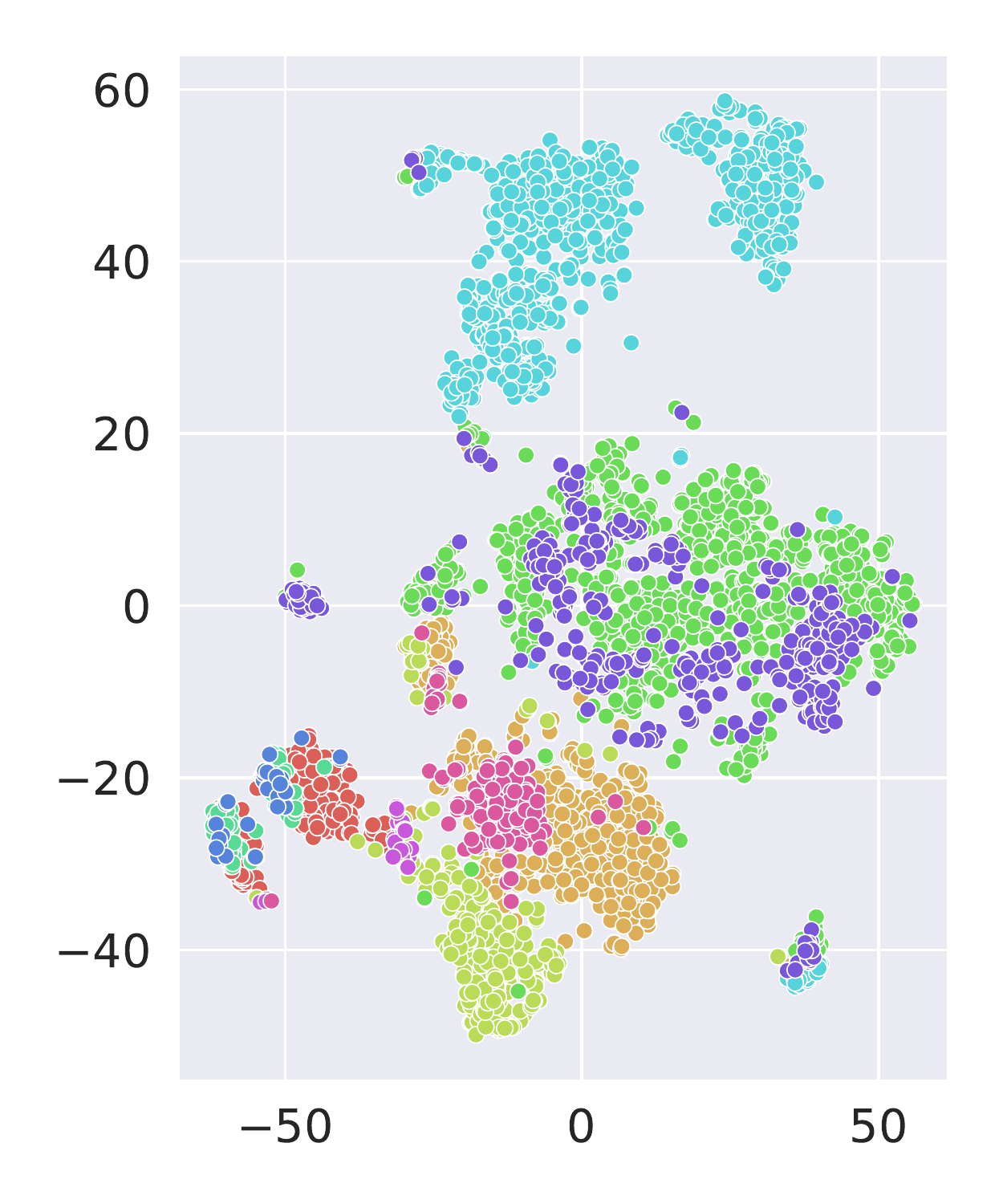}
}
\end{minipage}%
\begin{minipage}{0.33\linewidth}
\subfigure[Transformer: DB Index 3.59]{
\includegraphics[width=\linewidth]{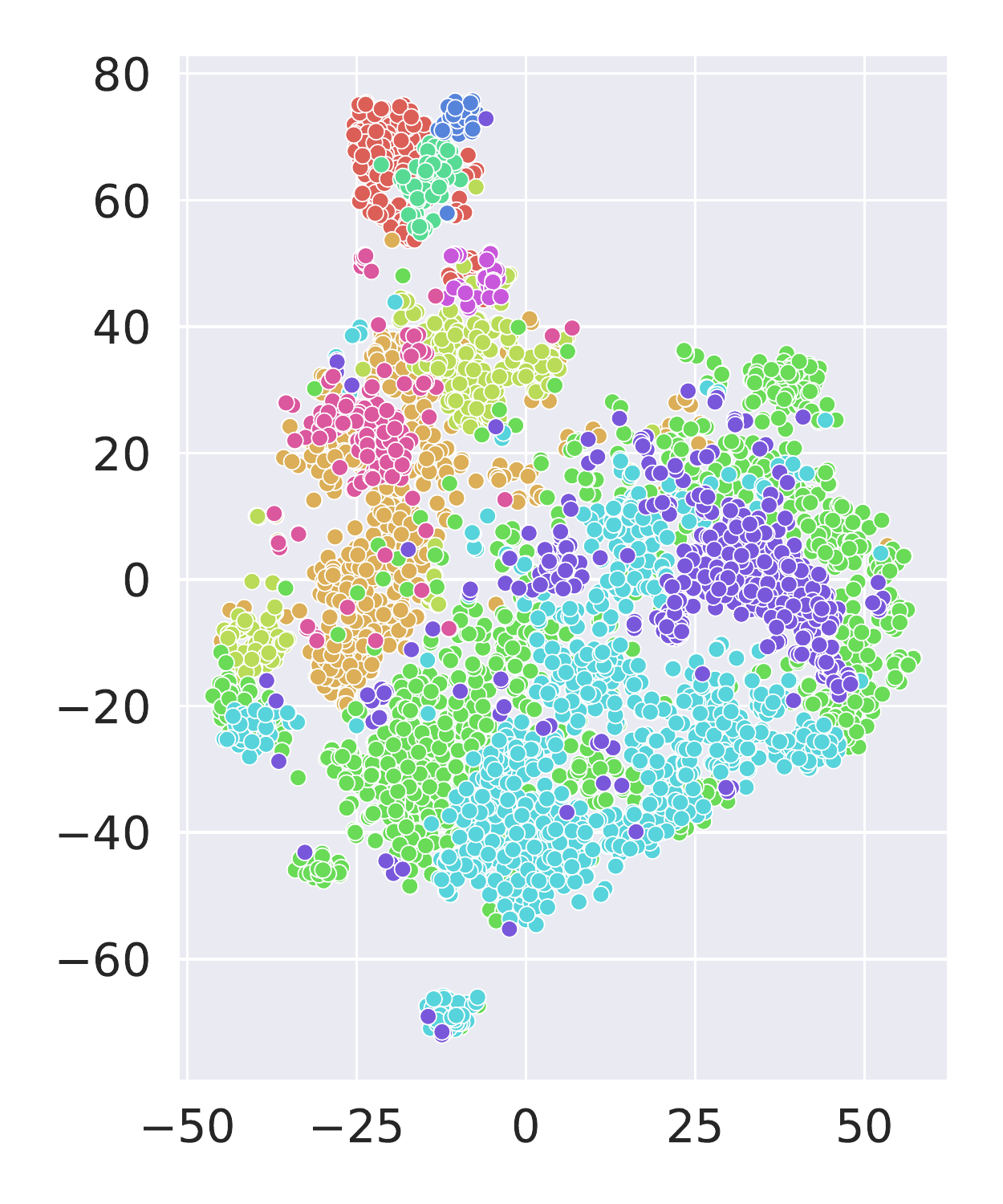}
}
\end{minipage}%
\begin{minipage}{0.33\linewidth}
\subfigure[\ourM{}: DB Index 2.19]{
\includegraphics[width=\linewidth]{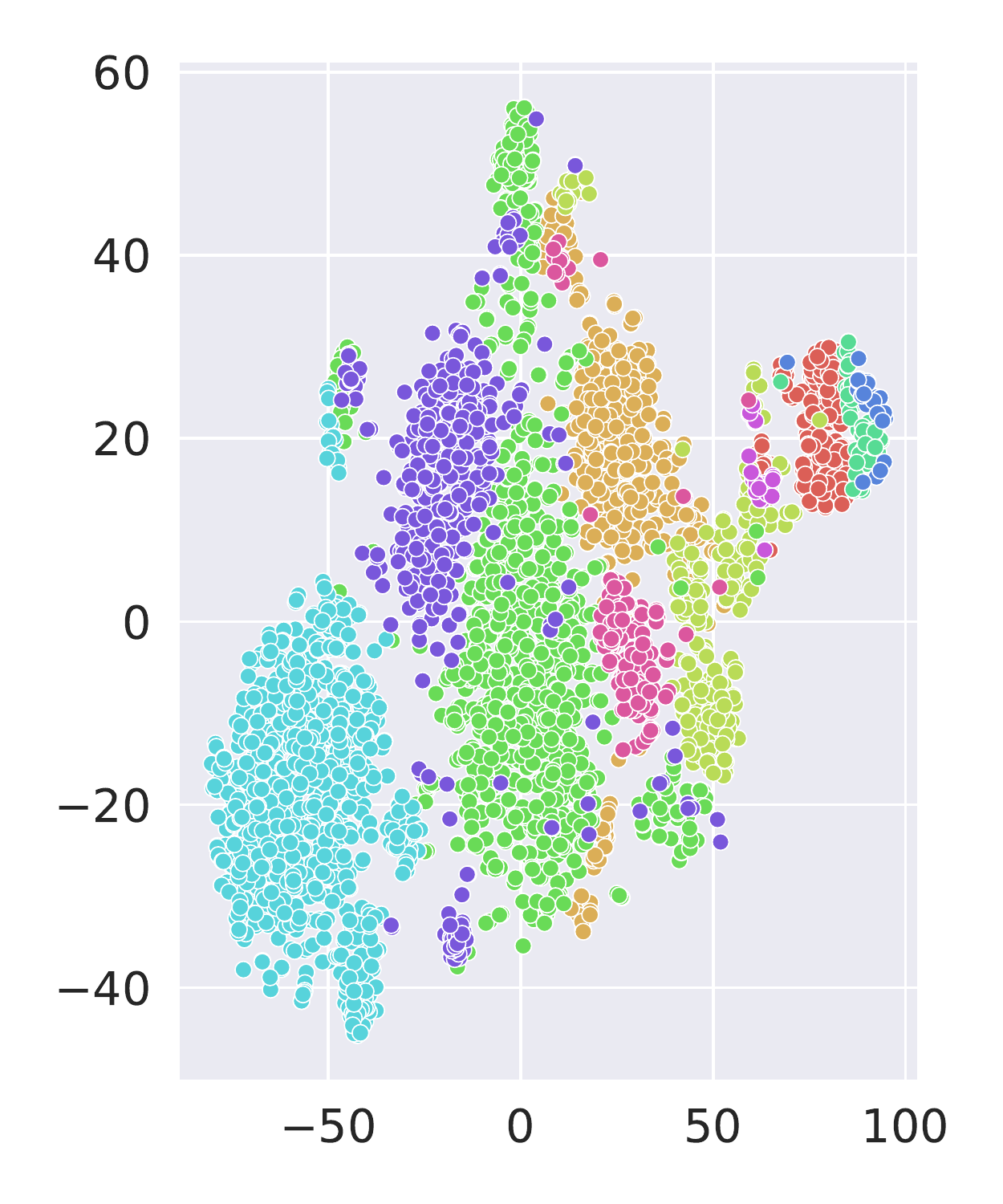}
}
\end{minipage}
\caption{Visualization of the representations learned by different models. Different colors indicate different scaffolds.}
\label{fig:viz_different_models_by_scaffold}
\end{figure}
We visualize the pre-trained representations to verify whether they capture the scaffold information~\cite{hu2016scaffold}, which is used to represent the core structures of bioactive compounds. Intuitively, molecules with the same scaffold share similar architectures, and therefore are expected to be close in the high-level representation space. Following \cite{li2020learn}, we choose ten representative scaffolds (denoted as $\mathcal{S}$)\footnote{We show the ten scaffolds in Appendix \ref{appendix:scaffold}. } and then randomly select $200k$ compounds.
For each compound whose scaffold lies in $\mathcal{S}$, we obtain its three representations: one from GNN pre-training with MLM only, one from Transformer pre-training with MLM only, and the third one from the  Transformer branch of our \ourM{} pre-training. We visualize the features through t-SNE~\cite{van2008visualizing} and the results are shown in Figure~\ref{fig:viz_different_models_by_scaffold}. We find that for the representations obtained by GNN pre-training and Transformer pre-training,   the molecules in different scaffolds  are overlapped and mixed together (e.g., the green and purple nodes for GNN; the green and light blue nodes for Transformer). In contrast, for the representations obtained by \ourM{} pre-training, the molecules in different scaffolds are well separated, which demonstrates that our \ourM{} better captures the scaffold information. Quantitatively, in terms of the DB index \citep{davies1979cluster} (the smaller, the better), which is a metric to evaluate the clustering results, \ourM{} clearly outperforms GNN pre-training and Transformer pre-training.



\section{Conclusions and future work}
In this work, we proposed \ourM{}, a novel molecule pre-training method that leverages dual views of molecules.
The core idea of \ourM{} is to maximize the consistency between the two representations extracted from two views, in addition to predicting masked tokens.
We achieve state-of-the-art results on seven molecular property prediction tasks from MoleculeNet and three retrosynthesis tasks. 
For future work, first, we will combine with stronger GNN models, e.g. \cite{rong2020self,li2020learn}. Second, as discussed in Section \ref{discussion_limitation}, it is interesting to design a pre-training method that dynamically determines which view to use for a specific molecule instead of going through both views, so as to improve training efficiency. 
Finally, compressing the pre-trained model is another interesting topic. 


{
\small{
\bibliographystyle{mybst}
\bibliography{mybib}

\begin{thebibliography}{59}
\providecommand{\natexlab}[1]{#1}
\providecommand{\url}[1]{\texttt{#1}}
\expandafter\ifx\csname urlstyle\endcsname\relax
  \providecommand{\doi}[1]{doi: #1}\else
  \providecommand{\doi}{doi: \begingroup \urlstyle{rm}\Url}\fi

\bibitem[Blum \& Mitchell(1998)Blum and Mitchell]{blum1998combining}
Avrim Blum and Tom Mitchell.
\newblock Combining labeled and unlabeled data with co-training.
\newblock In \emph{Proceedings of the eleventh annual conference on
  Computational learning theory}, pp.\  92--100, 1998.

\bibitem[Chen et~al.(2020)Chen, Kornblith, Norouzi, and Hinton]{chen2020simple}
Ting Chen, Simon Kornblith, Mohammad Norouzi, and Geoffrey Hinton.
\newblock A simple framework for contrastive learning of visual
  representations.
\newblock In \emph{International conference on machine learning}, pp.\
  1597--1607. PMLR, 2020.

\bibitem[Chithrananda et~al.(2020)Chithrananda, Grand, and
  Ramsundar]{chithrananda2020chemberta}
Seyone Chithrananda, Gabe Grand, and Bharath Ramsundar.
\newblock Chemberta: Large-scale self-supervised pretraining for molecular
  property prediction.
\newblock \emph{arXiv preprint arXiv:2010.09885}, 2020.

\bibitem[Coley et~al.(2017)Coley, Rogers, Green, and Jensen]{coley2017computer}
Connor~W Coley, Luke Rogers, William~H Green, and Klavs~F Jensen.
\newblock Computer-assisted retrosynthesis based on molecular similarity.
\newblock \emph{ACS central science}, 3\penalty0 (12):\penalty0 1237--1245,
  2017.

\bibitem[Cortes \& Vapnik(1995)Cortes and Vapnik]{cortes1995support}
Corinna Cortes and Vladimir Vapnik.
\newblock Support-vector networks.
\newblock \emph{Machine learning}, 20\penalty0 (3):\penalty0 273--297, 1995.

\bibitem[Dai et~al.(2019)Dai, Li, Coley, Dai, and Song]{GLN}
Hanjun Dai, Chengtao Li, Connor~W Coley, Bo~Dai, and Le~Song.
\newblock Retrosynthesis prediction with conditional graph logic network.
\newblock In \emph{NEURIPS}, 2019.

\bibitem[David et~al.(2020)David, Thakkar, Mercado, and
  Engkvist]{david2020molecular}
Laurianne David, Amol Thakkar, Roc{\'\i}o Mercado, and Ola Engkvist.
\newblock Molecular representations in ai-driven drug discovery: a review and
  practical guide.
\newblock \emph{Journal of Cheminformatics}, 12\penalty0 (1):\penalty0 1--22,
  2020.

\bibitem[Davies \& Bouldin(1979)Davies and Bouldin]{davies1979cluster}
David~L Davies and Donald~W Bouldin.
\newblock A cluster separation measure.
\newblock \emph{IEEE transactions on pattern analysis and machine
  intelligence}, \penalty0 (2):\penalty0 224--227, 1979.

\bibitem[Devlin et~al.(2018)Devlin, Chang, Lee, and Toutanova]{devlin2018bert}
Jacob Devlin, Ming-Wei Chang, Kenton Lee, and Kristina Toutanova.
\newblock Bert: Pre-training of deep bidirectional transformers for language
  understanding.
\newblock \emph{arXiv preprint arXiv:1810.04805}, 2018.

\bibitem[Dong et~al.(2018)Dong, Xu, and Xu]{8462506}
Linhao Dong, Shuang Xu, and Bo~Xu.
\newblock Speech-transformer: A no-recurrence sequence-to-sequence model for
  speech recognition.
\newblock In \emph{2018 IEEE International Conference on Acoustics, Speech and
  Signal Processing (ICASSP)}, pp.\  5884--5888, 2018.
\newblock \doi{10.1109/ICASSP.2018.8462506}.

\bibitem[Dosovitskiy et~al.(2021)Dosovitskiy, Beyer, Kolesnikov, Weissenborn,
  Zhai, Unterthiner, Dehghani, Minderer, Heigold, Gelly, Uszkoreit, and
  Houlsby]{dosovitskiy2021an}
Alexey Dosovitskiy, Lucas Beyer, Alexander Kolesnikov, Dirk Weissenborn,
  Xiaohua Zhai, Thomas Unterthiner, Mostafa Dehghani, Matthias Minderer, Georg
  Heigold, Sylvain Gelly, Jakob Uszkoreit, and Neil Houlsby.
\newblock An image is worth 16x16 words: Transformers for image recognition at
  scale.
\newblock In \emph{International Conference on Learning Representations}, 2021.
\newblock URL \url{https://openreview.net/forum?id=YicbFdNTTy}.

\bibitem[Elbayad et~al.(2019)Elbayad, Gu, Grave, and Auli]{elbayad2019depth}
Maha Elbayad, Jiatao Gu, Edouard Grave, and Michael Auli.
\newblock Depth-adaptive transformer.
\newblock \emph{arXiv preprint arXiv:1910.10073}, 2019.

\bibitem[Fabian et~al.(2020)Fabian, Edlich, Gaspar, Segler, Meyers, Fiscato,
  and Ahmed]{fabian2020molecular}
Benedek Fabian, Thomas Edlich, H{\'e}l{\'e}na Gaspar, Marwin Segler, Joshua
  Meyers, Marco Fiscato, and Mohamed Ahmed.
\newblock Molecular representation learning with language models and
  domain-relevant auxiliary tasks.
\newblock \emph{arXiv preprint arXiv:2011.13230}, 2020.

\bibitem[Grill et~al.(2020{\natexlab{a}})Grill, Strub, Altch\'{e}, Tallec,
  Richemond, Buchatskaya, Doersch, Avila~Pires, Guo, Gheshlaghi~Azar, Piot,
  kavukcuoglu, Munos, and Valko]{BYOL2020}
Jean-Bastien Grill, Florian Strub, Florent Altch\'{e}, Corentin Tallec, Pierre
  Richemond, Elena Buchatskaya, Carl Doersch, Bernardo Avila~Pires, Zhaohan
  Guo, Mohammad Gheshlaghi~Azar, Bilal Piot, koray kavukcuoglu, Remi Munos, and
  Michal Valko.
\newblock Bootstrap your own latent - a new approach to self-supervised
  learning.
\newblock In H.~Larochelle, M.~Ranzato, R.~Hadsell, M.~F. Balcan, and H.~Lin
  (eds.), \emph{Advances in Neural Information Processing Systems}, volume~33,
  pp.\  21271--21284. Curran Associates, Inc., 2020{\natexlab{a}}.
\newblock URL
  \url{https://proceedings.neurips.cc/paper/2020/file/f3ada80d5c4ee70142b17b8192b2958e-Paper.pdf}.

\bibitem[Grill et~al.(2020{\natexlab{b}})Grill, Strub, Altch{\'e}, Tallec,
  Richemond, Buchatskaya, Doersch, Pires, Guo, Azar,
  et~al.]{grill2020bootstrap}
Jean-Bastien Grill, Florian Strub, Florent Altch{\'e}, Corentin Tallec,
  Pierre~H Richemond, Elena Buchatskaya, Carl Doersch, Bernardo~Avila Pires,
  Zhaohan~Daniel Guo, Mohammad~Gheshlaghi Azar, et~al.
\newblock Bootstrap your own latent: A new approach to self-supervised
  learning.
\newblock \emph{arXiv preprint arXiv:2006.07733}, 2020{\natexlab{b}}.

\bibitem[Hadsell et~al.(2006)Hadsell, Chopra, and LeCun]{1640964}
R.~Hadsell, S.~Chopra, and Y.~LeCun.
\newblock Dimensionality reduction by learning an invariant mapping.
\newblock In \emph{2006 IEEE Computer Society Conference on Computer Vision and
  Pattern Recognition (CVPR'06)}, volume~2, pp.\  1735--1742, 2006.
\newblock \doi{10.1109/CVPR.2006.100}.

\bibitem[Hassani \& Khasahmadi(2020)Hassani and
  Khasahmadi]{hassani2020contrastive}
Kaveh Hassani and Amir~Hosein Khasahmadi.
\newblock Contrastive multi-view representation learning on graphs.
\newblock In \emph{International Conference on Machine Learning}, pp.\
  4116--4126. PMLR, 2020.

\bibitem[He et~al.(2021)He, Mattsson, Forsberg, Bjerrum, Engkvist, Tyrchan,
  Czechtizky, et~al.]{he2021transformer}
Jiazhen He, Felix Mattsson, Marcus Forsberg, Esben~Jannik Bjerrum, Ola
  Engkvist, Christian Tyrchan, Werngard Czechtizky, et~al.
\newblock Transformer neural network for structure constrained molecular
  optimization.
\newblock 2021.

\bibitem[He et~al.(2016)He, Zhang, Ren, and Sun]{he2016deep}
Kaiming He, Xiangyu Zhang, Shaoqing Ren, and Jian Sun.
\newblock Deep residual learning for image recognition.
\newblock In \emph{Proceedings of the IEEE conference on computer vision and
  pattern recognition}, pp.\  770--778, 2016.

\bibitem[He et~al.(2020)He, Fan, Wu, Xie, and Girshick]{he2020momentum}
Kaiming He, Haoqi Fan, Yuxin Wu, Saining Xie, and Ross Girshick.
\newblock Momentum contrast for unsupervised visual representation learning.
\newblock In \emph{Proceedings of the IEEE/CVF Conference on Computer Vision
  and Pattern Recognition}, pp.\  9729--9738, 2020.

\bibitem[Ho(1995)]{ho1995random}
Tin~Kam Ho.
\newblock Random decision forests.
\newblock In \emph{Proceedings of 3rd international conference on document
  analysis and recognition}, volume~1, pp.\  278--282. IEEE, 1995.

\bibitem[Honda et~al.(2019)Honda, Shi, and Ueda]{honda2019smiles}
Shion Honda, Shoi Shi, and Hiroki~R Ueda.
\newblock Smiles transformer: pre-trained molecular fingerprint for low data
  drug discovery.
\newblock \emph{arXiv preprint arXiv:1911.04738}, 2019.

\bibitem[Hu et~al.(2020)Hu, Liu, Gomes, Zitnik, Liang, Pande, and
  Leskovec]{hu2019strategies}
Weihua Hu, Bowen Liu, Joseph Gomes, Marinka Zitnik, Percy Liang, Vijay Pande,
  and Jure Leskovec.
\newblock Strategies for pre-training graph neural networks.
\newblock In \emph{International Conference on Learning Representations}, 2020.
\newblock URL \url{https://openreview.net/forum?id=HJlWWJSFDH}.

\bibitem[Hu et~al.(2016)Hu, Stumpfe, and Bajorath]{hu2016scaffold}
Ye~Hu, Dagmar Stumpfe, and Jürgen Bajorath.
\newblock Computational exploration of molecular scaffolds in medicinal
  chemistry.
\newblock \emph{Journal of Medicinal Chemistry}, 59\penalty0 (9):\penalty0
  4062--4076, 2016.
\newblock \doi{10.1021/acs.jmedchem.5b01746}.
\newblock URL \url{https://doi.org/10.1021/acs.jmedchem.5b01746}.
\newblock PMID: 26840095.

\bibitem[Huang \& Von~Lilienfeld(2016)Huang and
  Von~Lilienfeld]{huang2016communication}
Bing Huang and O~Anatole Von~Lilienfeld.
\newblock Communication: Understanding molecular representations in machine
  learning: The role of uniqueness and target similarity, 2016.

\bibitem[Ioffe \& Szegedy(2015)Ioffe and Szegedy]{ioffe2015batch}
Sergey Ioffe and Christian Szegedy.
\newblock Batch normalization: Accelerating deep network training by reducing
  internal covariate shift.
\newblock In \emph{International conference on machine learning}, pp.\
  448--456. PMLR, 2015.

\bibitem[Irwin \& Shoichet(2005)Irwin and Shoichet]{doi:10.1021/ci049714}
John~J. Irwin and Brian~K. Shoichet.
\newblock Zinc - a free database of commercially available compounds for
  virtual screening.
\newblock \emph{Journal of Chemical Information and Modeling}, 45\penalty0
  (1):\penalty0 177--182, 2005.

\bibitem[Kim et~al.(2020)Kim, Chen, Cheng, Gindulyte, He, He, Li, Shoemaker,
  Thiessen, Yu, Zaslavsky, Zhang, and Bolton]{10.1093/nar/gkaa971}
Sunghwan Kim, Jie Chen, Tiejun Cheng, Asta Gindulyte, Jia He, Siqian He,
  Qingliang Li, Benjamin~A Shoemaker, Paul~A Thiessen, Bo~Yu, Leonid Zaslavsky,
  Jian Zhang, and Evan~E Bolton.
\newblock {PubChem in 2021: new data content and improved web interfaces}.
\newblock \emph{Nucleic Acids Research}, 49\penalty0 (D1):\penalty0
  D1388--D1395, 11 2020.
\newblock ISSN 0305-1048.
\newblock \doi{10.1093/nar/gkaa971}.
\newblock URL \url{https://doi.org/10.1093/nar/gkaa971}.

\bibitem[Kingma \& Ba(2014)Kingma and Ba]{kingma2014adam}
Diederik~P Kingma and Jimmy Ba.
\newblock Adam: A method for stochastic optimization.
\newblock \emph{arXiv preprint arXiv:1412.6980}, 2014.

\bibitem[Kipf \& Welling(2016)Kipf and Welling]{kipf2016semi}
Thomas~N Kipf and Max Welling.
\newblock Semi-supervised classification with graph convolutional networks.
\newblock \emph{arXiv preprint arXiv:1609.02907}, 2016.

\bibitem[Li et~al.(2020{\natexlab{a}})Li, Xiong, Thabet, and
  Ghanem]{li2020deepergcn}
Guohao Li, Chenxin Xiong, Ali Thabet, and Bernard Ghanem.
\newblock Deepergcn: All you need to train deeper gcns.
\newblock \emph{arXiv preprint arXiv:2006.07739}, 2020{\natexlab{a}}.

\bibitem[Li et~al.(2020{\natexlab{b}})Li, Wang, Qiao, Chen, Yu, Yao, Gao, Xie,
  and Song]{li2020learn}
Pengyong Li, Jun Wang, Yixuan Qiao, Hao Chen, Yihuan Yu, Xiaojun Yao, Peng Gao,
  Guotong Xie, and Sen Song.
\newblock Learn molecular representations from large-scale unlabeled molecules
  for drug discovery.
\newblock \emph{arXiv preprint arXiv:2012.11175}, 2020{\natexlab{b}}.

\bibitem[Liu et~al.(2017)Liu, Ramsundar, Kawthekar, Shi, Gomes, Luu~Nguyen, Ho,
  Sloane, Wender, and Pande]{liu2017retrosynthetic}
Bowen Liu, Bharath Ramsundar, Prasad Kawthekar, Jade Shi, Joseph Gomes, Quang
  Luu~Nguyen, Stephen Ho, Jack Sloane, Paul Wender, and Vijay Pande.
\newblock Retrosynthetic reaction prediction using neural sequence-to-sequence
  models.
\newblock \emph{ACS central science}, 3\penalty0 (10):\penalty0 1103--1113,
  2017.

\bibitem[Liu et~al.(2019{\natexlab{a}})Liu, Demirel, and Liang]{liu2019ngram}
Shengchao Liu, Mehmet~F Demirel, and Yingyu Liang.
\newblock N-gram graph: Simple unsupervised representation for graphs, with
  applications to molecules.
\newblock In H.~Wallach, H.~Larochelle, A.~Beygelzimer, F.~d\textquotesingle
  Alch\'{e}-Buc, E.~Fox, and R.~Garnett (eds.), \emph{Advances in Neural
  Information Processing Systems}, volume~32. Curran Associates, Inc.,
  2019{\natexlab{a}}.
\newblock URL
  \url{https://proceedings.neurips.cc/paper/2019/file/2f3926f0a9613f3c3cc21d52a3cdb4d9-Paper.pdf}.

\bibitem[Liu et~al.(2019{\natexlab{b}})Liu, Ott, Goyal, Du, Joshi, Chen, Levy,
  Lewis, Zettlemoyer, and Stoyanov]{liu2019roberta}
Yinhan Liu, Myle Ott, Naman Goyal, Jingfei Du, Mandar Joshi, Danqi Chen, Omer
  Levy, Mike Lewis, Luke Zettlemoyer, and Veselin Stoyanov.
\newblock Roberta: A robustly optimized bert pretraining approach.
\newblock \emph{arXiv preprint arXiv:1907.11692}, 2019{\natexlab{b}}.

\bibitem[Lu et~al.(2019)Lu, Liu, Wang, Huang, Lin, and He]{lu2019molecular}
Chengqiang Lu, Qi~Liu, Chao Wang, Zhenya Huang, Peize Lin, and Lixin He.
\newblock Molecular property prediction: A multilevel quantum interactions
  modeling perspective.
\newblock In \emph{Proceedings of the AAAI Conference on Artificial
  Intelligence}, volume~33, pp.\  1052--1060, 2019.

\bibitem[Ma et~al.(2020)Ma, Bian, Rong, Huang, Xu, Xie, Ye, and
  Huang]{ma2020multi}
Hehuan Ma, Yatao Bian, Yu~Rong, Wenbing Huang, Tingyang Xu, Weiyang Xie, Geyan
  Ye, and Junzhou Huang.
\newblock Multi-view graph neural networks for molecular property prediction.
\newblock \emph{arXiv preprint arXiv:2005.13607}, 2020.

\bibitem[Maziarka et~al.(2020)Maziarka, Danel, Mucha, Rataj, Tabor, and
  Jastrz{\k{e}}bski]{maziarka2020molecule}
{\L}ukasz Maziarka, Tomasz Danel, S{\l}awomir Mucha, Krzysztof Rataj, Jacek
  Tabor, and Stanis{\l}aw Jastrz{\k{e}}bski.
\newblock Molecule attention transformer.
\newblock \emph{arXiv preprint arXiv:2002.08264}, 2020.

\bibitem[Mikolov et~al.(2013)Mikolov, Chen, Corrado, and
  Dean]{mikolov2013efficient}
Tomas Mikolov, Kai Chen, Greg Corrado, and Jeffrey Dean.
\newblock Efficient estimation of word representations in vector space.
\newblock \emph{arXiv preprint arXiv:1301.3781}, 2013.

\bibitem[Parmar et~al.(2018)Parmar, Vaswani, Uszkoreit, Kaiser, Shazeer, Ku,
  and Tran]{image2018imagetransformer}
Niki Parmar, Ashish Vaswani, Jakob Uszkoreit, Lukasz Kaiser, Noam Shazeer,
  Alexander Ku, and Dustin Tran.
\newblock Image transformer.
\newblock In Jennifer Dy and Andreas Krause (eds.), \emph{Proceedings of the
  35th International Conference on Machine Learning}, volume~80 of
  \emph{Proceedings of Machine Learning Research}, pp.\  4055--4064. PMLR,
  10--15 Jul 2018.
\newblock URL \url{http://proceedings.mlr.press/v80/parmar18a.html}.

\bibitem[Pesciullesi et~al.(2020)Pesciullesi, Schwaller, Laino, and
  Reymond]{pesciullesi2020transfer}
Giorgio Pesciullesi, Philippe Schwaller, Teodoro Laino, and Jean-Louis Reymond.
\newblock Transfer learning enables the molecular transformer to predict
  regio-and stereoselective reactions on carbohydrates.
\newblock \emph{Nature communications}, 11\penalty0 (1):\penalty0 1--8, 2020.

\bibitem[Rao et~al.(2019)Rao, Bhattacharya, Thomas, Duan, Chen, Canny, Abbeel,
  and Song]{rao2019evaluating}
Roshan Rao, Nicholas Bhattacharya, Neil Thomas, Yan Duan, Xi~Chen, John Canny,
  Pieter Abbeel, and Yun~S Song.
\newblock Evaluating protein transfer learning with tape.
\newblock \emph{Advances in Neural Information Processing Systems},
  32:\penalty0 9689, 2019.

\bibitem[Ren et~al.(2019)Ren, Ruan, Tan, Qin, Zhao, Zhao, and
  Liu]{NEURIPS2019_f63f65b5}
Yi~Ren, Yangjun Ruan, Xu~Tan, Tao Qin, Sheng Zhao, Zhou Zhao, and Tie-Yan Liu.
\newblock Fastspeech: Fast, robust and controllable text to speech.
\newblock In H.~Wallach, H.~Larochelle, A.~Beygelzimer, F.~d\textquotesingle
  Alch\'{e}-Buc, E.~Fox, and R.~Garnett (eds.), \emph{Advances in Neural
  Information Processing Systems}, volume~32, 2019.
\newblock URL
  \url{https://proceedings.neurips.cc/paper/2019/file/f63f65b503e22cb970527f23c9ad7db1-Paper.pdf}.

\bibitem[Rives et~al.(2021)Rives, Meier, Sercu, Goyal, Lin, Liu, Guo, Ott,
  Zitnick, Ma, and Fergus]{ESMpaper}
Alexander Rives, Joshua Meier, Tom Sercu, Siddharth Goyal, Zeming Lin, Jason
  Liu, Demi Guo, Myle Ott, C.~Lawrence Zitnick, Jerry Ma, and Rob Fergus.
\newblock Biological structure and function emerge from scaling unsupervised
  learning to 250 million protein sequences.
\newblock \emph{Proceedings of the National Academy of Sciences}, 118\penalty0
  (15), 2021.
\newblock ISSN 0027-8424.
\newblock \doi{10.1073/pnas.2016239118}.
\newblock URL \url{https://www.pnas.org/content/118/15/e2016239118}.

\bibitem[Rong et~al.(2020)Rong, Bian, Xu, Xie, Wei, Huang, and
  Huang]{rong2020self}
Yu~Rong, Yatao Bian, Tingyang Xu, Weiyang Xie, Ying Wei, Wenbing Huang, and
  Junzhou Huang.
\newblock Self-supervised graph transformer on large-scale molecular data.
\newblock \emph{Advances in Neural Information Processing Systems}, 33, 2020.

\bibitem[Schwaller et~al.(2019)Schwaller, Laino, Gaudin, Bolgar, Hunter, Bekas,
  and Lee]{schwaller2019molecular}
Philippe Schwaller, Teodoro Laino, Th{\'e}ophile Gaudin, Peter Bolgar,
  Christopher~A Hunter, Costas Bekas, and Alpha~A Lee.
\newblock Molecular transformer: a model for uncertainty-calibrated chemical
  reaction prediction.
\newblock \emph{ACS central science}, 5\penalty0 (9):\penalty0 1572--1583,
  2019.

\bibitem[Shen et~al.(2020)Shen, Liu, Wu, and Xie]{shen2020molgnn}
Xiaoke Shen, Yang Liu, You Wu, and Lei Xie.
\newblock Molgnn: Self-supervised motif learning graph neural network for drug
  discovery.
\newblock \emph{Machine Learning for Molecules Workshop at NeurIPS 2020}, 2020.
\newblock URL
  \url{https://ml4molecules.github.io/papers2020/ML4Molecules_2020_paper_4.pdf}.

\bibitem[Shi et~al.(2020)Shi, Xu, Guo, Zhang, and Tang]{G2Gs}
Chence Shi, Minkai Xu, Hongyu Guo, Ming Zhang, and Jian Tang.
\newblock A graph to graphs framework for retrosynthesis prediction.
\newblock In \emph{International Conference on Machine Learning}, pp.\
  8818--8827. PMLR, 2020.

\bibitem[Tetko et~al.(2020)Tetko, Karpov, Van~Deursen, and
  Godin]{tetko2020state}
Igor~V Tetko, Pavel Karpov, Ruud Van~Deursen, and Guillaume Godin.
\newblock State-of-the-art augmented nlp transformer models for direct and
  single-step retrosynthesis.
\newblock \emph{Nature communications}, 11\penalty0 (1):\penalty0 1--11, 2020.

\bibitem[Van~der Maaten \& Hinton(2008)Van~der Maaten and
  Hinton]{van2008visualizing}
Laurens Van~der Maaten and Geoffrey Hinton.
\newblock Visualizing data using t-sne.
\newblock \emph{Journal of machine learning research}, 9\penalty0 (11), 2008.

\bibitem[Vaswani et~al.(2017)Vaswani, Shazeer, Parmar, Uszkoreit, Jones, Gomez,
  Kaiser, and Polosukhin]{vaswani2017attention}
Ashish Vaswani, Noam Shazeer, Niki Parmar, Jakob Uszkoreit, Llion Jones,
  Aidan~N Gomez, Lukasz Kaiser, and Illia Polosukhin.
\newblock Attention is all you need.
\newblock \emph{arXiv preprint arXiv:1706.03762}, 2017.

\bibitem[Wang et~al.(2019)Wang, Guo, Wang, Sun, and Huang]{wang2019smiles}
Sheng Wang, Yuzhi Guo, Yuhong Wang, Hongmao Sun, and Junzhou Huang.
\newblock Smiles-bert: large scale unsupervised pre-training for molecular
  property prediction.
\newblock In \emph{Proceedings of the 10th ACM international conference on
  bioinformatics, computational biology and health informatics}, pp.\
  429--436, 2019.

\bibitem[Wang et~al.(2021)Wang, Wang, Cao, and Farimani]{wang2021molclr}
Yuyang Wang, Jianren Wang, Zhonglin Cao, and Amir~Barati Farimani.
\newblock Molclr: Molecular contrastive learning of representations via graph
  neural networks.
\newblock \emph{arXiv preprint arXiv:2102.10056}, 2021.

\bibitem[Weininger(1988)]{weininger1988smiles}
David Weininger.
\newblock Smiles, a chemical language and information system. 1. introduction
  to methodology and encoding rules.
\newblock \emph{Journal of chemical information and computer sciences},
  28\penalty0 (1):\penalty0 31--36, 1988.

\bibitem[Wu et~al.(2018)Wu, Ramsundar, Feinberg, Gomes, Geniesse, Pappu,
  Leswing, and Pande]{wu2018moleculenet}
Zhenqin Wu, Bharath Ramsundar, Evan~N Feinberg, Joseph Gomes, Caleb Geniesse,
  Aneesh~S Pappu, Karl Leswing, and Vijay Pande.
\newblock Moleculenet: a benchmark for molecular machine learning.
\newblock \emph{Chemical science}, 9\penalty0 (2):\penalty0 513--530, 2018.

\bibitem[Yan et~al.(2020)Yan, Ding, Zhao, Zheng, YANG, Yu, and
  Huang]{NEURIPS2020_819f46e5}
Chaochao Yan, Qianggang Ding, Peilin Zhao, Shuangjia Zheng, JINYU YANG, Yang
  Yu, and Junzhou Huang.
\newblock Retroxpert: Decompose retrosynthesis prediction like a chemist.
\newblock In H.~Larochelle, M.~Ranzato, R.~Hadsell, M.~F. Balcan, and H.~Lin
  (eds.), \emph{Advances in Neural Information Processing Systems}, volume~33,
  pp.\  11248--11258. Curran Associates, Inc., 2020.
\newblock URL
  \url{https://proceedings.neurips.cc/paper/2020/file/819f46e52c25763a55cc642422644317-Paper.pdf}.

\bibitem[Yang et~al.(2019)Yang, Swanson, Jin, Coley, Eiden, Gao, Guzman-Perez,
  Hopper, Kelley, Mathea, et~al.]{yang2019analyzing}
Kevin Yang, Kyle Swanson, Wengong Jin, Connor Coley, Philipp Eiden, Hua Gao,
  Angel Guzman-Perez, Timothy Hopper, Brian Kelley, Miriam Mathea, et~al.
\newblock Analyzing learned molecular representations for property prediction.
\newblock \emph{Journal of chemical information and modeling}, 59\penalty0
  (8):\penalty0 3370--3388, 2019.

\bibitem[Zhu et~al.(2020)Zhu, Xia, Wu, He, Qin, Zhou, Li, and
  Liu]{zhu2020incorporating}
Jinhua Zhu, Yingce Xia, Lijun Wu, Di~He, Tao Qin, Wengang Zhou, Houqiang Li,
  and Tie-Yan Liu.
\newblock Incorporating bert into neural machine translation.
\newblock \emph{arXiv preprint arXiv:2002.06823}, 2020.

\bibitem[Zhu et~al.(2021)Zhu, Wu, Xia, Xie, Qin, Zhou, Li, and Liu]{zhu2021iot}
Jinhua Zhu, Lijun Wu, Yingce Xia, Shufang Xie, Tao Qin, Wengang Zhou, Houqiang
  Li, and Tie-Yan Liu.
\newblock {\{}IOT{\}}: Instance-wise layer reordering for transformer
  structures.
\newblock In \emph{International Conference on Learning Representations}, 2021.
\newblock URL \url{https://openreview.net/forum?id=ipUPfYxWZvM}.

\end{thebibliography}
}}

\appendix
\section{Experiment Setup}
\subsection{Detailed configuration of the GNN branch}\label{appendix:deepergcn}
The GNN branch in \ourM{} is a variant of DeeperGCN \citep{li2020deepergcn}, which is stacked by 12 identical blocks. Each block consists of a batch normalization layer, a nonlinear layer and a graph convolutional layer sequentially, with a skip  connection connected the input and the output. In each graph convolutional layer, each node will fuse its neighbor edge representations and its neighbor node representations with an aggregation layer. Specifically, in \ourM{}, the aggregation layer is a concatenation of maximize, minimize and average pooling. 
\subsection{Finetuning hyperparameters}\label{appendix:finetuninghyperparaments}
We summarize the finetuning hyperparameters in Table \ref{tab:finetuning_hyperparameter}.
\begin{table}[!h]
    \centering
    \begin{tabular}{lcc}
    \toprule
    \textbf{Hyperparam} & \textbf{Classification} & \textbf{Regression} \\
    \midrule
      Learning rate &$\{5e-5, 1e-4, 2e-4\}$  & $\{5e-5, 1e-4, 2e-4\}$ \\
      Batch Size & $\{ 8,16,32\}$& $\{ 8,16,32\}$ \\
      Weight Decay & $\{0.1,0.01\}$& $\{0.1,0.01\}$\\
      Max Epochs &$10$&$10$ \\
      Learning Rate Decay &Linear&Linear \\
      Warmup ratio& $0.06$&$0.06$ \\
      Dropout &$0.1$ & $\{0.1, 0.2, 0.3\}$\\
      \bottomrule
    \end{tabular}
    \caption{Hyperparameters for finetuning \ourM{}.}
    \label{tab:finetuning_hyperparameter}
\end{table}
\subsection{Scaffolds}\label{appendix:scaffold}
\begin{figure}[!htb]
    \centering
    \includegraphics[width=\textwidth]{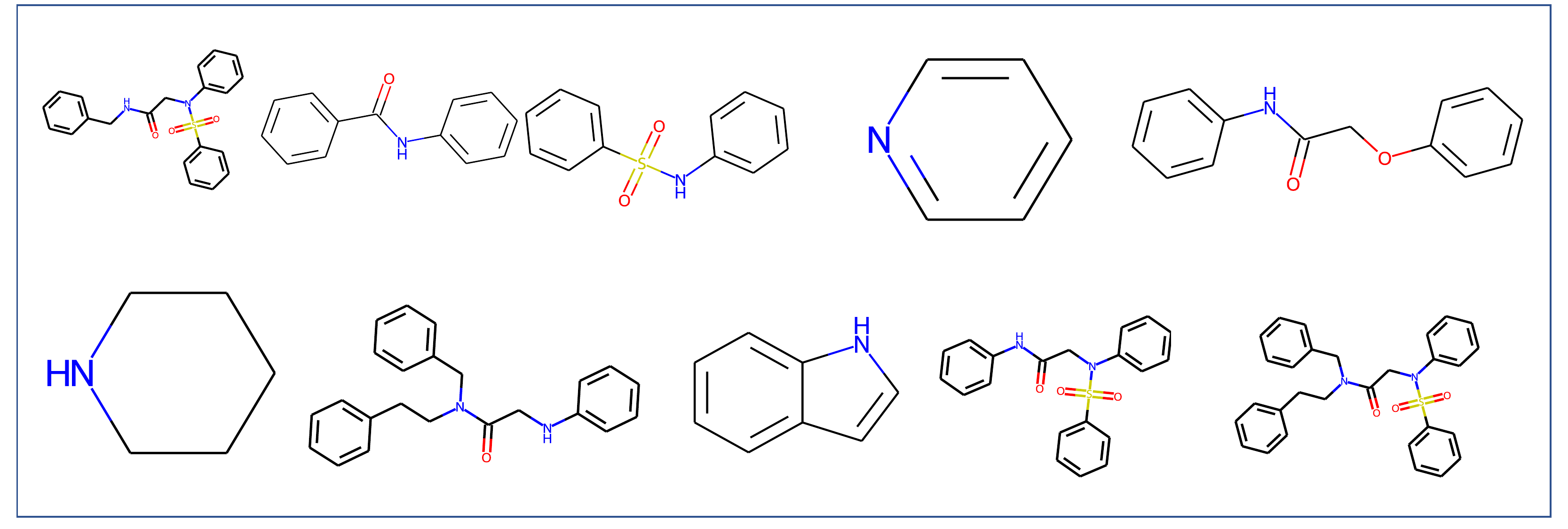}
    \caption{Ten scaffolds used in our visualization.}
    \label{fig:scaffold}
\end{figure}

In Figure \ref{fig:scaffold}, we present the scaffolds used in Figure \ref{fig:viz_different_models_by_scaffold} of the main paper. The scaffolds are extracted using the command \texttt{rdkit.Chem.Scaffolds.MurckoScaffold.MurckoScaffoldSmiles} of the RDKit package.
\section{More experimental results}

\subsection{Comparison with ChemBERTa \citep{chithrananda2020chemberta}}\label{appendix:chemberta}
\begin{table}[!htb]
    \centering
    \begin{tabular}{lcc}
    \toprule
    & BBBP &HIV \\
    \midrule
      ChemBERTa &$0.643$  &$0.622$  \\
      TF (MLM) & $0.749$ & $0.802$ \\
      \ourM{}$\Rightarrow$ TF &$0.781$  &$0.810$ \\
      \bottomrule
    \end{tabular}
    \caption{Comparison with ChemBERTa.}
    \label{tab:comwithchemberta}
\end{table}
We compare our method with ChemBERTa \citep{chithrananda2020chemberta} on two classification datasets, BBBP and HIV, which is also used in \citep{chithrananda2020chemberta}.
The results are summarized in Table \ref{tab:comwithchemberta}, where TF (MLM) means that we pre-train a Transformer model using masked language modeling only, and \ourM{}$\Rightarrow$TF means that we pre-train a Transformer using \ourM{}. TF (MLM) is an enhanced implementation of ChemBERTa, where we train models with more GPUs and larger batch size, and eventually obtain better results.
We can see that our proposed method  outperforms both ChemBERTa and TF (MLM), which shows the effectiveness of our method.


\subsection{Comparison with contrastive learning}\label{appendix:contrastive} Considering dual-view loss is related to contrastive loss, we also explore a corresponding variant: given a molecule $m$, the corresponding features from Transformer branch and GNN branch are denoted as $f_s(m)$ and $f_g(m)$ respectively. Given several other molecules $m_1,m_2,\cdots,m_n$, $f_g(m)$ should be similar to $f_s(m)$ while dissimilar to $f_s(m_1),\cdots,f_s(m_n)$, and $f_s(m)$ should be dissimilar to $f_g(m_1),\cdots,f_g(m_n)$. Such a variant achieves $77.1\%$ and $94.5\%$ ROC-AUC on BBBP and ClinTox dataset, respectively, and performs a little worse than \ourM{} ($78.1\%$ and $95.0\%$).

\subsection{Retrosynthesis}\label{appendix:retrosys}
We compare the results on retrosysthesis of existed works and our method in Table \ref{appendix:retrosys_more}. The results of Retrosim and neuralsym are from~\cite{GLN}.
On the three dataset, our method achieves the best result. 
\begin{table}[!htb]
\small
    \centering
    \begin{tabular}{lcccccc}
    \toprule
         \multirow{2}{*}{Methods} & \multicolumn{6}{c}{Top-$k$ accuracy (\%)}  \\
         \cmidrule{2-7}
         & $1$& $3$ & $5$& $10$&$20$ & $50$\\
         \midrule
         \multicolumn{7}{c}{Reaction types unknown on USPTO-50K}\\
         \midrule
         Transformer & $42.3$ & $61.9$ & $67.5$ & $72.9$ & $75.5$ & $77.1$ \\
         Pre-trained model as Encoder& $39.6$&$55.3$&$59.1$&$63.2$&$66.0$&$68.6$\\
          ChemBERTa fusion \cite{chithrananda2020chemberta}&$43.9$&$62.2$&$68.0$&$73.1$&$75.4$&$77.0$\\
         \ourM{} fusion &$46.1$&$65.2$&$70.4$&$74.3$&$76.1$&$77.5$ \\
          GLN w/ DeeperGCN & $52.1$ & $68.8$&$76.1$&$84.0$&$89.2$&$92.4$\\
         GLN w/ \ourM{}  &$54.2$&$70.5$&$77.2$&$84.9$&$90.0$&$92.7$\\
         \midrule
         RetroSim & $37.3$ &$54.7$& $63.3$&$74.1$&$82.0$&$85.3$ \\
         NeuralSym & $44.4$&$65.3$&$72.4$&$78.9$&$82.2$&$83.1$ \\
         GLN~\cite{GLN}&$52.6$&$68.0$&$75.1$&$83.1$&$88.5$&$92.1$\\
         \midrule
         \multicolumn{7}{c}{Reaction types give as prior on USPTO-50K}\\
         \midrule
         Transformer & $54.2$ & $73.6$ & $78.3$&$81.3$& $83.1$&$84.3$ \\
          ChemBERTa fusion \cite{chithrananda2020chemberta}&$56.4$&$74.7$&$78.9$&$81.8$&$83.3$&$84.5$\\
          \ourM{} fusion  &$57.5$&$75.5$&$80.2$&$83.1$&$84.2$&$85.1$\\
        GLN w/ DeeperGCN & $63.2$ & $79.2$&$85.0$&$90.0$&$92.0$&$93.1$\\
GLN w/ \ourM{}  &$66.5$&$81.2$&$86.6$&$90.5$&$92.8$&$93.5$\\
         \midrule
         RetroSim & $52.9$ &$73.8$& $81.2$&$88.1$&$91.8$&$92.9$ \\
         NeuralSym & $55.3$&$76.0$&$81.4$&$85.1$&$86.5$&$86.9$ \\
         GLN~\cite{GLN}&$63.2$&$77.5$&$83.4$&$89.1$&$92.1$&$93.2$\\
         \midrule
         \multicolumn{7}{c}{Retrosynthesis results on USPTO-full}\\
         \midrule
         Transformer & $42.9$ & $58.0$ & $62.4$ & $66.8$ &$69.8$& $72.5$ \\
          \ourM{} fusion &$45.0$&$59.6$&$63.9$&$67.9$&$70.7$&$73.2$ \\
         \midrule
         Retrosim & $32.8$ & -- & -- & $56.1$ \\
         neuralsym & $35.8$ & -- & -- & $60.8$ \\
         GLN~\cite{GLN} & $39.3$ & -- & -- & $63.7$\\
         \bottomrule
    \end{tabular}
    \caption{Results of top-$k$ exact match accuracy on Retrosynthesis. }
    \label{appendix:retrosys_more}
\end{table}


\end{document}